\documentstyle[12pt]{article}

\skewchar\fivmi='177
\skewchar\sixmi='177
\skewchar\sevmi='177
\skewchar\egtmi='177
\skewchar\ninmi='177
\skewchar\tenmi='177
\skewchar\elvmi='177
\skewchar\twlmi='177
\skewchar\frtnmi='177
\skewchar\svtnmi='177
\skewchar\twtymi='177
\def\@magscale#1{ scaled \magstep #1}
\font\twfvmi  = ammi10   \@magscale5 
\skewchar\twfvmi='177
\skewchar\fivsy='60
\skewchar\sixsy='60
\skewchar\sevsy='60
\skewchar\egtsy='60
\skewchar\ninsy='60
\skewchar\tensy='60
\skewchar\elvsy='60
\skewchar\twlsy='60
\skewchar\frtnsy='60
\skewchar\svtnsy='60
\skewchar\twtysy='60
\font\twfvsy  = amsy10   \@magscale5 
\skewchar\twfvsy='60

\catcode`@=11
\def\un#1{\relax\ifmmode\@@underline#1\else
        $\@@underline{\hbox{#1}}$\relax\fi}
\catcode`@=12

\def\a{\alpha}

\def\k{\kappa}
\def\l{\lambda}
\def\m{\mu}
\def\n{\nu}
\def\o{\omega}

\font\sc=font005                        
\font\ooo=circle10                      
\font\ro=manfnt                         
\def\kcl{{\hbox{\ro 6}}}                
\def\kcr{{\hbox{\ro 7}}}                
\def\ktl{{\hbox{\ro \char'134}}}        
\def\ktr{{\hbox{\ro \char'135}}}        
\def\kbl{{\hbox{\ro \char'136}}}        
\def\kbr{{\hbox{\ro \char'137}}}        

\def\bo{{\raise.15ex\hbox{\large$\Box$}}}               
\def\TH{{\raise.2ex\hbox{$\displaystyle \bigodot$}\mskip-4.7mu \llap H \;}}
\def\face{{\raise.2ex\hbox{$\displaystyle \bigodot$}\mskip-2.2mu \llap {$\ddot
        \smile$}}}                                      

\def\sp#1{{}^{#1}}                              
\def\leftrightarrowfill{$\mathsurround=0pt \mathord\leftarrow \mkern-6mu
        \cleaders\hbox{$\mkern-2mu \mathord- \mkern-2mu$}\hfill
        \mkern-6mu \mathord\rightarrow$}
\def\dvec#1{\vbox{\ialign{##\crcr
        \leftrightarrowfill\crcr\noalign{\kern-1pt\nointerlineskip}
        $\hfil\displaystyle{#1}\hfil$\crcr}}}           

\def\frac#1#2{{\textstyle{#1\over\vphantom2\smash{\raise.20ex
        \hbox{$\scriptstyle{#2}$}}}}}                   
\def\sfrac#1#2{{\vphantom1\smash{\lower.5ex\hbox{\small$#1$}}\over
        \vphantom1\smash{\raise.4ex\hbox{\small$#2$}}}} 
\def\bfrac#1#2{{\vphantom1\smash{\lower.5ex\hbox{$#1$}}\over
        \vphantom1\smash{\raise.3ex\hbox{$#2$}}}}       
\def\afrac#1#2{{\vphantom1\smash{\lower.5ex\hbox{$#1$}}\over#2}}    

\newskip\humongous \humongous=0pt plus 1000pt minus 1000pt
\def\caja{\mathsurround=0pt}

\newif\ifdtup
\def\panorama{\global\dtuptrue \openup2\jot \caja
        \everycr{\noalign{\ifdtup \global\dtupfalse
        \vskip-\lineskiplimit \vskip\normallineskiplimit
        \else \penalty\interdisplaylinepenalty \fi}}}
\def\eqalignno#1{\panorama \tabskip=\humongous                  
        \halign to\displaywidth{\hfil$\displaystyle{##}$
        \tabskip=0pt&$\displaystyle{{}##}$\hfil
        \tabskip=\humongous&\llap{$##$}\tabskip=0pt
        \crcr#1\crcr}}

\def\ref#1{$\sp{#1)}$}

\parskip=\medskipamount                 
\thicklines                         

\def\oldheadpic{                                
        \setlength{\unitlength}{.4mm}
        \thinlines
        \par
        \begin{picture}(349,16)
        \put(325,16){\line(1,0){4}}
        \put(330,16){\line(1,0){4}}
        \put(340,16){\line(1,0){4}}
        \put(335,0){\line(1,0){4}}
        \put(340,0){\line(1,0){4}}
        \put(345,0){\line(1,0){4}}
        \put(329,0){\line(0,1){16}}
        \put(330,0){\line(0,1){16}}
        \put(339,0){\line(0,1){16}}
        \put(340,0){\line(0,1){16}}
        \put(344,0){\line(0,1){16}}
        \put(345,0){\line(0,1){16}}
        \put(329,16){\oval(8,32)[bl]}
        \put(330,16){\oval(8,32)[br]}
        \put(339,0){\oval(8,32)[tl]}
        \put(345,0){\oval(8,32)[tr]}
        \end{picture}
        \par
        \thicklines
        \vskip.2in}
\def\oldtitle#1#2#3#4{\oldheadpic\begin{center}\vglue.5in{\large\bf #1}\\[.6in]
        {#2}\\[.1in] {\it Department of Physics and Astronomy}\\
        {\it University of Maryland, College Park, MD 20742}\\[.6in]
        Physics Publication \#{#3}\\ {#4}\\[1.5in] {\bf ABSTRACT}\\[.1in]
        \end{center} \begin{quotation}}                 
\def\oldTitle#1#2#3#4#5#6#7{\oldheadpic\begin{center} \vglue .4in
        {\large\bf #1}\\[.4in]
        {#2}\\[.1in] {\it Department of Physics and Astronomy}\\
        {\it University of Maryland, College Park, MD 20742}\\[.1in]
        {#3}\\[.1in] {\it {#4}}\\ {\it {#5}}\\[.4in]
        Physics Publication \#{#6}\\ {#7}\\[.5in] {\bf ABSTRACT}\\[.1in]
        \end{center} \begin{quotation}}                 
\def\border{                                            
        \setlength{\unitlength}{1mm}
        \newcount\xco
        \newcount\yco
        \xco=-24
        \yco=12
        \begin{picture}(140,0)
        \put(\xco,\yco){$\ktl$}
        \advance\yco by-1
        {\loop
        \put(\xco,\yco){$\kcl$}
        \advance\yco by-2
        \ifnum\yco>-240
        \repeat
        \put(\xco,\yco){$\kbl$}}
        \xco=158
        \yco=12
        \put(\xco,\yco){$\ktr$}
        \advance\yco by-1
        {\loop
        \put(\xco,\yco){$\kcr$}
        \advance\yco by-2
        \ifnum\yco>-240
        \repeat
        \put(\xco,\yco){$\kbr$}}
        \put(-20,11){\tiny University of Maryland Elementary Particle
Physics University of Maryland Elementary Particle Physics University of
Maryland Elementary Particle Physics}
        \put(-20,-241.5){\tiny University of Maryland Elementary
Particle Physics University of Maryland Elementary Particle Physics
University of Maryland Elementary Particle Physics}
        \end{picture}
        \par\vskip-8mm}
\def\bordero{                                           
        \setlength{\unitlength}{1mm}
        \newcount\xco
        \newcount\yco
        \xco=-24
        \yco=12
        \begin{picture}(140,0)
        \put(\xco,\yco){$\ktl$}
        \advance\yco by-1
        {\loop
        \put(\xco,\yco){$\kcl$}
        \advance\yco by-2
        \ifnum\yco>-240
        \repeat
        \put(\xco,\yco){$\kbl$}}
        \xco=158
        \yco=12
        \put(\xco,\yco){$\ktr$}
        \advance\yco by-1
        {\loop
        \put(\xco,\yco){$\kcr$}
        \advance\yco by-2
        \ifnum\yco>-240
        \repeat
        \put(\xco,\yco){$\kbr$}}
        \put(-20,12){\ooo
bacdefghidfghghdhededbihdgdfdfhhdheidhdhebaaahjhhdahba

hgdedge
   hgfdiehhgdigicba}
        \put(-20,-241.5){\ooo
ababaighefdbfghgeahgdfgafagihdidihiidhiagfedhadbfd

ecdcdfa
   gdcbhaddhbgfchbgfdacfediacbabab}
        \end{picture}
        \par\vskip-8mm}
\def\headpic{                                           
        \indent
        \setlength{\unitlength}{.4mm}
        \thinlines
        \par
        \begin{picture}(29,16)
        \put(165,16){\line(1,0){4}}
        \put(170,16){\line(1,0){4}}
        \put(180,16){\line(1,0){4}}
        \put(175,0){\line(1,0){4}}
        \put(180,0){\line(1,0){4}}
        \put(185,0){\line(1,0){4}}
        \put(169,0){\line(0,1){16}}
        \put(170,0){\line(0,1){16}}
        \put(179,0){\line(0,1){16}}
        \put(180,0){\line(0,1){16}}
        \put(184,0){\line(0,1){16}}
        \put(185,0){\line(0,1){16}}
        \put(169,16){\oval(8,32)[bl]}
        \put(170,16){\oval(8,32)[br]}
        \put(179,0){\oval(8,32)[tl]}
        \put(185,0){\oval(8,32)[tr]}
        \end{picture}
        \par\vskip-6.5mm
        \thicklines}
\def\title#1#2#3#4{\border\headpic {\hbox to\hsize{#4 \hfill UMDEPP #3}}\par
        \begin{center} \vglue .5in {\large\bf #1}\\[.6in]
        {#2}\\[.1in] {\it Department of Physics and Astronomy}\\
        {\it University of Maryland, College Park, MD 20742}\\[1.5in]
        {\bf ABSTRACT}\\[.1in] \end{center} \begin{quotation}}  
\def\Title#1#2#3#4#5#6#7{\border\headpic
        {\hbox to\hsize{#7 \hfill UMDEPP #6}}\par
        \begin{center} \vglue .4in {\large\bf #1}\\[.4in]
        {#2}\\[.1in] {\it Department of Physics and Astronomy}\\
        {\it University of Maryland, College Park, MD 20742}\\[.1in]
        {#3}\\[.1in] {\it {#4}}\\ {\it {#5}}\\[.5in] {\bf ABSTRACT}\\[.1in]
        \end{center} \begin{quotation}}                 
\def\endtitle{\end{quotation}\newpage}                  

\def\sect#1{\bigskip\medskip \goodbreak \noindent{\bf {#1}} \nobreak \medskip}
\def\refs{\sect{REFERENCES} \footnotesize \frenchspacing \parskip=0pt}
\def\Item{\par\hang\textindent}

\begin{document}

%

\def\gfrac#1#2{\frac {\scriptstyle{#1}}
        {\mbox{\raisebox{-.6ex}{$\scriptstyle{#2}$}}}}
\def\gg{{\hbox{\sc g}}}
\border\headpic {\hbox to\hsize{
November 1991 \hfill {UMDEPP 92-110}}}\par
\bigskip
\bigskip
\bigskip
{\hfill {\it Submitted to Physical Review Letters}}
\begin{center}
\vglue .4in
{\large\bf The Instability of String--Theoretic Black Holes}
\\[.6in]
Gerald Gilbert\footnote {Research supported in part by NSF grant
\# PHY--91-41926}
\footnote {Bitnet: gng@umdhep ; Internet: gng@umdhep.umd.edu}\\[.3in]
{\it Department of Physics and Astronomy\\
University of Maryland at College Park\\
College Park, MD 20742-4111 USA}\\[.8in]

{\bf ABSTRACT}\\[.3in]
\end{center}
\begin{quotation}

{It is demonstrated that static, charged, spherically--symmetric
black holes in string theory are classically and
catastrophically unstable to linearized
perturbations in four dimensions, and moreover that unstable
modes appear for arbitrarily small positive values of the charge.
This catastrophic classical instability dominates and is distinct
from much smaller and less significant effects such as
possible quantum mechanical evaporation.
The classical instability of the string--theoretic black hole
contrasts sharply with the situation which obtains for the
Reissner--Nordstr\"om black hole of general relativity, which has
been shown by Chandrasekhar to be perfectly
stable to linearized perturbations at the event horizon.}

\endtitle

\sect{{\bf SECTION ONE}}

\noindent{In [1] the perturbations of string--theoretic
black holes were
investigated by deriving and performing an analysis of the equations
which result upon perturbing and
linearizing about the background, in the most general possible way,
putative black hole solutions to the string equations
of motion.\footnote{The definitions and conventions
employed in this article conform to those used in [1].}
The results of that detailed calculation
revealed that the potential which surrounds the
string--theoretic black hole
as a result of linearized incident perturbations differs radically
from the potential which surrounds the Reissner--Nordstr\"om black hole
of general relativity. Here we will employ that result to ascertain
that the static, charged string--theoretic black hole is classically
and catastrophically unstable in four dimensions. Such a classical
instability dominates much smaller quantum mechanical effects such as
possible quantum mechanical evaporation.}

\noindent{The charged string--theoretic black hole we will consider
was first derived in [2]. It was later obtained independently in [3]
by solving the
equations of motion which arise from the string--inspired effective action
characterizing the heterotic string at low energies. The
black hole is a solution
at the level of the Born approximation in string theory, i.e.,
string--loop effects are not incorporated in the solution, and it
is also approximate in the $\a'$--expansion of the sigma--model defined
on the world sheet of the string.
The black hole field configuration so obtained is specified
by the metric tensor corresponding to the invariant squared
line--element given by}

$$ds^2=-e^{2f_0}dt^2+e^{2f_1}d\varphi^2+e^{2f_2}dr^2+e^{2f_3}d\theta^2
{}~,\eqno(1)$$

\noindent{where the metric functions are determined to be}

$$e^{2f_0}=e^{-2f_2}=1-{2M\over r} ~,\eqno(2)$$

$$e^{2f_3}=r\left(r-{Q^2e^{2\Phi_0}\over M}\right) ~,\eqno(3)$$

\noindent{and}

$$e^{2f_1}=e^{2f_3}\sin^2\theta ~,\eqno(4)$$

\noindent{along with the field strengths $\Phi=\Phi(r)$
for a massless dilaton and ${\cal F}_{\m\n}$ for a U(1)
gauge field given by:}

$$e^{-2\Phi}=e^{-2\Phi_0}-{Q^2\over Mr} ~,\eqno(5)$$

\noindent{and}

$${\cal F}=-Qe^{-2f_3}~dt\wedge dr ~,\eqno(6)$$

\noindent{where $M$ is the mass of the black hole, $Q$ is the
charge of the gauge field and $\Phi_0$ is the
constant, asymptotic value of the dilaton field. It will be observed
that this solution is spherically--symmetric and asymptotically--flat.}

\sect{{\bf SECTION TWO}}

\noindent{In [1] a detailed calculation of invigorating
albeit unavoidably lugubrious
prolixity was performed in which it was demonstrated that the
equations for the linearized axial perturbations of the
charged string--theoretic black hole are given by}

$$\left({d^2\over dr_*^2}+\o^2\right){\cal Z}_{1,2} = V_{1,2}{\cal
Z}_{1,2} ~,\eqno(7)$$

\noindent{where the perturbation potentials $V_{1,2}$ were explicitly
calculated and are recorded for convenience
in the appendix to this paper. Here ${\cal Z}_{1,2}$,
the two independent axial modes of the perturbed black hole,
are comprised of linear combinations of incident electromagnetic and
gravitational perturbations $Y_1$ and $Y_2$, where the
relation between the two is given by}

$${\cal Z}_1 = u_2Y_1-u_1Y_2 ~,\eqno(8)$$

\noindent{and}

$${\cal Z}_2 = v_1Y_2-v_2Y_1 ~.\eqno(9)$$

\noindent{In the above equations $\pmatrix{u_1\cr u_2\cr}$
and $\pmatrix{v_1\cr v_2\cr}$
are the two independent eigenvectors, associated
with the eigenvalues $V_1$ and $V_2$, respectively,
belonging to the matrix}

$$M = \k\pmatrix{A&C\cr C&B\cr} ~,\eqno(10)$$

\noindent{where}

$$\k\equiv e^{2f_0-2f_3} ~,\eqno(11)$$

$$A\equiv n^2+2-e^{-2f_0+2f_3}{\cal T}_1+4Q^2e^{-2\Phi}e^{-2f_3} ~,
\eqno(12)$$

$$B\equiv n^2-e^{-2f_0+2f_3}{\cal T}_2 ~,\eqno(13)$$

\noindent{and}

$$C\equiv 2nQe^{-\Phi}e^{-f_3} ~,\eqno(14)$$

\noindent{and we have also defined}

$${\cal T}_1\equiv {d^2\over dr_*^{~2}}\left(\Phi-f_0-f_3\right)-
\left[{d\over dr_*}\left(\Phi-f_0-f_3\right)\right]^2 ~,\eqno(15)$$

\noindent{and}

$${\cal T}_2\equiv {d^2\over dr_*^{~2}}f_3-\left({d\over dr_*}f_3
\right)^2 ~.\eqno(16)$$

\noindent{Here $n^2 = (l-1)(l+2)$, where $l$ is the integer--valued
index of the Gegenbauer polynomial $C_{l+2}^{-3/2}(\theta)$ which
was shown in [1] to completely describe the angular dependence of the
axial perturbations. The perturbations are characterized by a constant,
non--dispersive frequency $\o$ with time--dependence given by
$e^{i\o t}$. In the preceeding equations the luminosity coordinate
distance $r_*$ has been introduced,
which for this problem is given by $r_*\equiv r+2M\log\left({r\over 2M}
-1\right)$.\footnote{Note that this differs from the definition of $r_*$
appropriate to the analysis of the Reissner--Nordstr\"om black hole.}
We note that one may verify that,
in the double--limit $Q\rightarrow 0$ and $\Phi_0\rightarrow 0$,
eq.(7) governing the evolution of ${\cal Z}_2$ reduces {\it exactly} to
the equation for the (single) axial perturbation of the Schwarzschild
black hole.\footnote{See, e.g., [4] for an analysis of the perturbations
of the Schwarzschild black hole.} We may also note that in the limit
$\Phi_0\rightarrow 0$ eq.'s(7) do not (and should not) reduce to the
pair of axial perturbation equations which describe the
Reissner--Nordstr\"om black hole. (This is a consequence of the presence
in the effective action of a term linear in the dilaton field proportional
to $\Phi {\cal F}^2$, implicitly contained within the expansion of
$e^{-2\Phi}{\cal F}^2$, which dictates that a non--constant dilaton field
must appear if ${\cal F}\not= 0$ in the string--theoretic solution. As
a result the Reissner--Nordstr\"om black hole does not even approximately
solve the string equations of motion.)}

\noindent{As explained in [1], there are no axial modes associated with
the perturbation in the dilaton field, and thus there are only two
axial perturbation potentials. In contrast, as discussed
in [1], there are
three perturbation potentials associated with the polar modes of the
perturbed black hole, arising from electromagnetic, gravitational and
dilatonic contributions. It is very important to note that
the equations for the polar and axial perturbations are entirely
distinct and completely free of mixing terms. As a consequence
of this a necessary condition for the stability of the black hole is that
the configuration be stable to the incidence of {\it both} axial and
polar waves. We will henceforth focus on the implications of the axial
perturbations since their derivation and analysis is less involved
than that required for the polar modes and is sufficient to demonstrate
instabilty. The essential point of interest with respect to the
question of classical stability is that one of the two
axial potentials, $V_2$, {\it assumes negative values}
in the regime of weak string--coupling. (We may note that
the derivation in [3] of the black hole solution is demonstrably legitimate
only in the regime of weak coupling.)\footnote{The intrinsic
coupling parameter $g_s$ which measures the strength of
string interactions is related to the
dilaton field through the equation $g_s^2=e^{2\Phi_0}.$} The
dependence of the perturbation potentials upon $\Phi_0$ is of
course expected since the scale--invariance of the classical string
action is violated in the solutions to the equations by the appearance
in the black hole configuration of the dimensionful parameters $Q$ and $M$.}

\noindent{The explicit expressions for the
perturbation potentials derived in [1] are sufficiently
complicated to warrant numerical investigation.\footnote{The computer
program {\it Mathematica} [5] is adequate for this purpose.}
This analysis reveals that the perturbed
configuration, as a consequence of external waves impinging on the
string--theoretic black hole, is characterized by a
short--range, one--dimensional negative potential
well which surrounds the black hole just outside the event horizon.
As a consequence of this we will find that eq.(7) will allow
a finite number of discrete, non--degenerate bound states.
In sharp contrast to this behavior the perturbation potentials of the
Reissner--Nordstr\"om black hole are strictly positive outside of the
event horizon, which furnishes a sufficient condition for the stability
of that solution [6].}

\noindent{To proceed we note that eq.(7) can be rearranged as}

$${\partial^2{\cal Z}_{1,2}\over
\partial t^2} = \left({d^2\over
dr_*^2}-V_{1,2}\right){\cal Z}_{1,2} ~,\eqno(17)$$

\noindent{where we have implicitly
restored the time--dependent factor $e^{i\o t}$.
If we introduce the definition ${\cal A}_i\equiv -{d^2\over dr_*^2}+V_i$
($i=1,2$) the operator of interest is then given by}

$${\cal A}_2\equiv -{d^2\over dr_*^2}+V_2 ~.\eqno(18)$$

\noindent{In [7] elliptic operators of the
form of ${\cal A}$ were analyzed. (It is important in the analysis
that the perturbation potential $V$ is bounded, which may be readily
verified for both $V_1$ and $V_2$.) It was proved that if an operator
of the form of ${\cal A}$ fails to be strictly positive on the
Hilbert space $L^2({\bf M})$ when
acting on a manifold ${\bf R}\times {\bf M}$, where
${d\over dr_*}$ is the derivative operator associated with a complete
Riemannian metric on ${\bf M}$, there will exist smooth initial data
which generate an unstable solution to eq.(17). Such a solution will
grow unboundedly with time.
It is straightforward to verify that the elliptic
operator ${\cal A}_2$ is {\it not} strictly positive in the region
outside of the event horizon of the black hole, as a consequence of
which we may deduce that $\o^2$ is negative. This may be done in a number
of ways. A conclusive but extremely tedious
demonstration of this follows upon explicitly applying ${d^2\over dr_*^2}$
to the independent mode ${\cal Z}_2$ after deriving $Y_1$ and $Y_2$ as
Fredholm series (as explained in [1]), making
use of eq.(9). This calculation reveals
that there are positive values of ${d^2{\cal Z}_2\over dr_*^2}$
corresponding to negative values of $V_2$, and
thus $-{d^2\over dr_*^2}+V_2={\cal A}_2<0$. We thus learn from the fact
that the operator ${\cal A}_2$ fails to be strictly positive
that the black hole is fundamentally unstable to linearized
classical perturbations. Further numerical investigation reveals that
as the value of the charge of the black hole is reduced to arbitrarily
small, positive values the unstable modes persist if the value of
$\Phi_0$ is also reduced.}

\sect{{\bf SECTION THREE}}

\noindent{The classical effect we have discovered leads to the
rapid destabilization of the string--theoretic black hole,
and dominates over smaller effects such as possible quantum
mechanical evaporation. The classical instability is significant
in view of the fact that, as remarked above, the general relativistic
black hole which is the ``analogue" of this configuration,
the Reissner--Nordstr\"om
solution, is perfectly stable to classical linearized perturbations.}

\noindent{As pointed out in [3] and mentioned above, due to the linear
coupling of the dilaton to ${\cal F}^2$ the Reissner--Nordstr\"om
solution of general relativity is not even an approximate solution
of string theory. Thus,
the fact that the string--intrinsic black hole considered in this article
is inherently unstable may have profound consequences, as it uniquely
appears in
a more fundamental theory than general relativity. The implications of
this result have yet to be fully explored. In particular, it is
important to understand fully the unstable evolution of the black hole.
Work is in progress on this and related issues [8,9].}

\noindent{{\it Acknowledgements:} The author wishes to thank D. Brill,
S. Giddings, J. Hartle, M. Perry, J. Schwarz, A. Strominger, R. Wald
and N. Warner for comments.}

\sect{{\bf APPENDIX}}

\noindent{The explicit expressions for the potentials, associated
with the axial perturbations, which surround the
electrically--charged string--theoretic black holes are given by [1]:}

$$V_1(r,Q,M,\Phi_0)=\l\left(v
+{\sqrt {\Delta}}\right) ~,\eqno({\rm A}1)$$

\noindent{and}

$$V_2(r,Q,M,\Phi_0)=\l\left(v
-{\sqrt {\Delta}}\right) ~,\eqno({\rm A}2)$$

\noindent{where:}

$$\l(r,Q,M,\Phi_0)=\left[-8e^{2\Phi_0}r^4\left(Mr-
Q^2e^{2\Phi_0}\right)^2\right]^{-1} ~,\eqno({\rm A}3)$$

$$\eqalignno{v(r,Q,M,\Phi_0)=&-32M^2Q^4e^{2\Phi_0}
-40M^2Q^4e^{6\Phi_0}+(32M^3Q^2+16 MQ^4e^{2\Phi_0}\cr
&+64M^3Q^2e^{4\Phi_0}+24MQ^4e^{6\Phi_0})r-
(16M^2Q^2+36M^4e^{2\Phi_0}\cr
&+60M^2Q^2e^{4\Phi_0} +
3Q^4e^{6\Phi_0}+16M^2Q^2e^{4\Phi_0}n^2)r^2 +
(48M^3e^{2\Phi_0}\cr &+16MQ^2e^{4\Phi_0}
+16M^3e^{2\Phi_0}n^2+8MQ^2e^{4\Phi_0}n^2)r^3
-(16 M^2e^{2\Phi_0}\cr &+8M^2e^{2\Phi_0}n^2)r^4 ~,&({\rm A}4)\cr}$$

\noindent{and}

$$\eqalignno{\Delta(r,Q,M,\Phi_0)=&1024M^4Q^8e^{4\Phi_0}\cr &+
(-2048M^5Q^6e^{2\Phi_0} -
1024M^3Q^8e^{4\Phi_0}+3072M^5Q^6e^{6\Phi_0}
+512M^3Q^8e^{8\Phi_0}) r\cr
&+(1024M^6 Q^4+2048M^4Q^6e^{2\Phi_0}-
6912M^6Q^4e^{4\Phi_0}+256M^2Q^8e^{4\Phi_0}\cr &-
3840M^4Q^6e^{6\Phi_0}+2304M^6Q^4e^{8\Phi_0}-
448M^2Q^8e^{8\Phi_0}+768 M^4 Q^6e^{10\Phi_0}\cr
&+64M^2Q^8e^{12\Phi_0}+1024M^4Q^6e^{6\Phi_0}n^2)r^2\cr &+
(-1024M^5Q^4+3840M^7Q^2e^{2\Phi_0}-512M^3Q^6e^{2\Phi_0}+
7296M^5Q^4e^{4\Phi_0}\cr &-5760M^7Q^2e^{6\Phi_0}+
1344M^3Q^6e^{6\Phi_0}-3648M^5Q^4e^{8\Phi_0}+96MQ^8e^{8\Phi_0}\cr &-
736M^3Q^6e^{10\Phi_0}-48MQ^8e^{12\Phi_0}-
2048M^5Q^4e^{4\Phi_0}n^2-1024M^3Q^6e^{6\Phi_0}n^2)r^3\cr
&+(256M^4Q^4-3968M^6Q^2e^{2\Phi_0}
+3600M^8e^{4\Phi_0}-1920M^4Q^4e^{4\Phi_0}\cr &+6432M^6Q^2e^{6\Phi_0}
-96 M^2Q^6e^{6\Phi_0}+
1656M^4Q^4e^{8\Phi_0}+ 168 M^2Q^6e^{10\Phi_0}\cr &+9Q^8e^{12\Phi_0}+
1024M^6Q^2e^{2\Phi_0}n^2+2048M^4Q^4e^{4\Phi_0}n^2+
256M^2Q^6e^{6\Phi_0}n^2)r^4\cr &+
(1024M^5Q^2e^{2\Phi_0}-3840M^7e^{4\Phi_0}-1792M^5Q^2e^{6\Phi_0}
-192 M^3Q^4e^{8\Phi_0}\cr &-1024M^5Q^2e^{2\Phi_0}n^2-
512 M^3Q^4e^{4\Phi_0}n^2)r^5\cr &+
(1024M^6e^{4\Phi_0}+256M^4Q^2e^{2\Phi_0}n^2)r^6 ~,&({\rm A}5)\cr}$$

\noindent{where $n^2=(l-1)(l+2)$ and $l$ is an integer.}

\refs

\Item{[1]} G. Gilbert, {\it On the Perturbations of String--Theoretic
Black Holes}, UMDEPP--92--094 (Revised Version), November 1991.
\Item{[2]} G.W. Gibbons and K. Maeda, {\it Nucl. Phys.} {\bf B298}
(1988) 741.
\Item{[3]} D. Garfinkle, G.T. Horowitz and A. Strominger, {\it Phys.
Rev.} {\bf D43} (1991) 3140.
\Item{[4]} S. Chandrasekhar, {\it Proc. Roy. Soc.} {\bf A343} (1975)
289.
\Item{[5]} S. Wolfram, {\it Mathematica}, Addison--Wesley, 1988.
\Item{[6]} S. Chandrasekhar, {\it Proc. Roy. Soc.} {\bf A365} (1979)
453.
\Item{[7]} R. Wald, {\it On the Instability of the $n$$=$$1$
Einstein--Yang--Mills Black Holes and Mathematically Related Systems},
Enrico Fermi Institute preprint, 1991.
\Item{[8]} S. Giddings and G. Gilbert, in preparation.
\Item{[9]} G. Gilbert, in preparation.

\end{document}